\documentclass[aps,prl,twocolumn]{revtex4}
\usepackage{graphicx}
\usepackage{dcolumn}
\usepackage{bm}

\begin{document}

\preprint{}

\title{Spin-polarized Mn$^{2+}$ emission from Manganese-doped colloidal nanocrystals}

\author{R. Viswanatha$^1$, J. M. Pietryga$^1$, V. I. Klimov$^1$, S. A. Crooker$^2$}

\affiliation{$^1$Chemistry Division, Los Alamos National Laboratory, Los Alamos, NM 87545}
\affiliation{$^2$National High Magnetic Field Laboratory, Los Alamos, NM 87545}

\date{\today}
\begin{abstract}
We report magneto-photoluminescence studies of strongly quantum-confined ``0-D" diluted magnetic semiconductors (DMS), realized in Mn$^{2+}$-doped ZnSe/CdSe core/shell colloidal nanocrystals. In marked contrast to their 3-D (bulk), 2-D (quantum well), 1-D (quantum wire), and 0-D (self-assembled quantum dot) DMS counterparts, the ubiquitous yellow emission band from internal \emph{d-d} ($^4T_1 \rightarrow ^6A_1$) transitions of the Mn$^{2+}$ ions in these nanocrystals is \emph{not} suppressed in applied magnetic fields and \emph{does} become circularly polarized. This polarization tracks the Mn$^{2+}$ magnetization, and is accompanied by a sizable energy splitting between right- and left-circular emission components that scales with the exciton-Mn \emph{sp-d} coupling strength (which, in turn, is tunable with nanocrystal size).  These data highlight the influence of strong quantum confinement on both the excitation and the emission mechanisms of magnetic ions in DMS nano-materials.

\end{abstract}
\maketitle
The coupling between band electrons and local magnetic moments underpins many fascinating phenomena in condensed matter physics. In semiconductors, these couplings are simply realized and often studied in diluted magnetic semiconductors (DMS) \cite{Furdyna, Becker, Book}, where the strong \emph{sp-d} exchange interactions between electron/hole spins and the embedded magnetic atoms (typically Mn) can lead, for example, to giant exciton \emph{g}-factors, magnetic polarons, or carrier-mediated ferromagnetism in both II-VI and III-V materials. To address how \emph{sp-d} interactions are affected -- and potentially controlled -- by quantum confinement and wavefunction engineering, advances in molecular-beam epitaxy (MBE) and colloidal synthesis have focused considerable attention on DMS systems of reduced dimensionality: 2-D quantum wells, 1-D wires, and 0-D epitaxial dots and (even smaller) nanocrystals. Recent studies indicate substantial influence arising from even \emph{single} Mn atoms \cite{Besombes, Hoffman, Norris, Hawrylak}, and suggest that the \emph{sp-d} exchange `constants' themselves can be modified \cite{Merkulov, Stern, Bussian}.

A defining feature of wide-gap A$^{\textmd{II}}_{1-x}$Mn$_x$B$^{\textmd{VI}}$ DMSs ($E_{g}$ $\agt$2.2~eV) is a prominent yellow photoluminescence (PL) band centered at $\sim$2.1 eV, that originates from internal ($^4T_1 \rightarrow ^6A_1$) transitions within excited $3d^5$ shells of the Mn$^{2+}$ ions \cite{Furdyna, Becker}. That is, a band-edge exciton can either recombine radiatively with rate $k_R$, \emph{or} transfer its energy to a Mn$^{2+}$ ion via an Auger-like process (described below) with rate $k_{Mn}$ that depends on the exciton-Mn coupling. It is known that at low temperatures and in magnetic fields, this Mn$^{2+}$ PL band is suppressed and remains unpolarized, while the band-edge exciton PL increases and circularly polarizes. This universal behavior has been observed in DMS crystals, epilayers, quantum wells, quantum wires, and in epitaxial quantum dots \cite{Furdyna, Becker, LeeRamdas, Kim, Lee, Falk, Oh, Shibata, Hundt, Chernenko}. To date however, the magneto-optical and polarization properties of this Mn$^{2+}$ PL have not been reported in colloidal DMS nanocrystals, despite widespread technological interest in their use as efficient phosphors \cite{Ratna} or spintronic materials \cite{Beaulac} and despite being materials that evince the strongest quantum confinement and potentially largest exciton-Mn coupling.

Here we measure the magneto-PL properties of Mn$^{2+}$-doped ZnSe/CdSe core/shell colloidal nanocrystals. In stark contrast to their 3-D (bulk), 2-D (quantum well), 1-D (wires and rods) and even their 0-D (epitaxial quantum dot) DMS counterparts, the yellow Mn$^{2+}$ PL band in these nanocrystals is \emph{not} suppressed in applied magnetic fields and \emph{does} develop a sizable circular polarization. This polarization tracks the magnetization of the embedded Mn$^{2+}$ atoms, and is accompanied by an unexpectedly large energy splitting between right- and left-circular PL components that scales with the overall strength of the \emph{sp-d} interaction (which, in turn, we tune with CdSe shell thickness). These data highlight the critical role of strong quantum confinement on both the excitation and the emission of Mn$^{2+}$ atoms in DMS nano-materials.

Mn$^{2+}$-doped ZnSe/CdSe core/shell nanocrystals (NCs) were grown using colloidal methods \cite{Norris, Ivanov}. All ZnSe cores have 17~\AA~radii. Five doping levels yielded cores averaging $\langle n_{Mn}\rangle$=0.8, 1.6, 2.6, 5.2, and 9.6 Mn$^{2+}$ ions/core (average Mn concentration up to $\sim$2\%), as determined by inductively-coupled plasma optical emission spectroscopy.  Nonmagnetic (undoped) cores were also grown. Ensembles of cores were then overcoated with CdSe shells of thickness 0-6($\pm$2)~\AA, which reduces quantum confinement and lowers the NC band-edge from 3 eV to $\sim$2.4 eV. Nonresonant magneto-PL was measured using weak 3.06 or 3.81 eV laser excitation (100 $\mu$W/cm$^2$) in the Faraday geometry (\textbf{B}$\parallel$\textbf{k}), with dilute NC films mounted in the variable-temperature insert of a 8~T magnet. Importantly, magnetic circular dichroism (MCD) spectroscopy was also performed on all films to measure the Zeeman splitting of the \emph{1S} (band-edge) exciton absorption peak. All Mn-doped NCs show enhanced Zeeman splittings that follow Brillouin ($B_{5/2}$) functions, indicating strong \emph{sp-d} coupling of the bands to the paramagnetic, spin-5/2 Mn$^{2+}$ ions \cite{Norris, Beaulac, Bussian}. This provides an independent measure of the Mn$^{2+}$ magnetization within a given NC sample and allows to compare the overall strength of the \emph{sp-d} interaction between samples. Quantum confinement and \emph{sp-d} coupling fall rapidly with increasing shell thickness as previously observed \cite{Bussian}, however these NCs do not exhibit \emph{sp-d} inversion since the NC band-edges did not fall below $\sim$2.4 eV.

\begin{figure}[tbp]
\includegraphics[width=.49\textwidth]{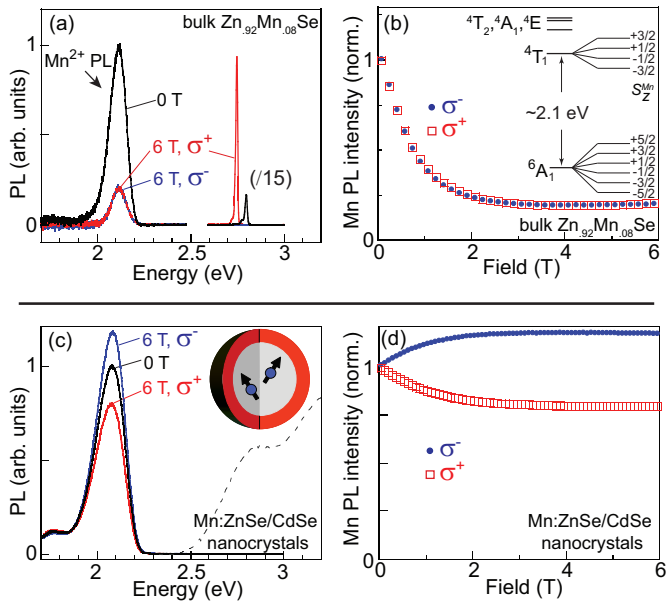}
\caption{(a,b) Magneto-PL from bulk ZnMnSe at 4~K, showing conventional DMS behavior. The yellow $^4T_1$$\rightarrow$$^6A_1$ Mn$^{2+}$ PL at $\sim$2.1~eV is suppressed by magnetic fields and remains unpolarized, while 2.8~eV exciton PL (scaled down 15$\times$) increases. Similar behavior is found in 3D, 2D, 1D, and 0D DMS grown by MBE. (c,d) The contrasting magneto-PL from Mn:ZnSe/CdSe nanocrystals ($\langle n_{Mn}\rangle$=2.6 Mn/core; $T$=1.8~K). The Mn$^{2+}$ PL is \emph{not} suppressed by magnetic fields, and \emph{does} develop a sizable circular polarization (and, exciton PL never appears). The dashed line shows the NC absorption.}
\label{fig1}
\end{figure}

To most clearly present the new aspects of the Mn$^{2+}$ PL from these NCs, we first show, by way of comparison, the characteristic magneto-PL from a traditional wide-gap II-VI DMS.  Fig. 1(a) shows PL from a Zn$_{.92}$Mn$_{.08}$Se epilayer grown by MBE.  At 0~T, the broad Mn$^{2+}$ PL band is clearly visible at $\sim$2.1~eV, as is the PL from exciton recombination at the 2.8~eV band-edge.  With applied magnetic field the Mn$^{2+}$ PL is rapidly and equally suppressed in both $\sigma^\pm$ circular polarizations (Fig. 1b), while the exciton PL increases many-fold and becomes completely $\sigma^+$ polarized. We observed similar behavior in a variety of other DMS epilayers and quantum wells.

This behavior -- a suppression of the unpolarized Mn$^{2+}$ PL and an enhancement and polarization of the exciton PL -- is a universal characteristic of all non-resonant magneto-PL studies of DMS materials reported to date, aspects of which have been reported in bulk crystals \cite{Furdyna, Becker, LeeRamdas, Kim, Lee}, in quantum wells \cite{Falk}, quantum wires \cite{Oh}, and in ``self-assembled" epitaxial quantum dots \cite{Shibata, Hundt, Chernenko}. Although the precise mechanism of energy transfer from excitons to the Mn$^{2+}$ $3d^5$ shell is still debated \cite{Nawrocki, Falk, Lee, Chernenko}, its marked field dependence indicates a \emph{spin-dependent} excitation transfer as detailed by Nawrocki \cite{Nawrocki} and later refined by Chernenko \cite{Chernenko}.  The relevant selection rules require that energy transfer conserves the spin projection of the Mn+exciton system along \textbf{B}, $S^{Mn}_z + s^{ex}_z$. Using standard notation \cite{Furdyna, Becker}, the spin-aligned ground state of the half-filled Mn$^{2+}$ $3d^5$ shell ($^6A_1$) has total spin $S^{Mn}$=5/2, whilst the lowest crystal-field-split excited states ($^4T_1$, $^4T_2$, $^4A_1$, $^4E$) have one flipped spin and therefore $S^{Mn}$=3/2. Applied fields Zeeman-split these levels [inset, Fig. 1(b)]. ``Bright" excitons (with \emph{total} spin+orbital angular momentum projection $J_z=\pm 1$) have $s^{ex}_z=0$, and therefore they \emph{cannot} excite Mn$^{2+}$ out of $S^{Mn}_z$=$\pm$5/2 levels. However, the -5/2 level becomes predominantly populated as the paramagnetic Mn$^{2+}$ polarize in applied magnetic fields. Concurrently, in the Faraday geometry, excitons rapidly populate the lowest $J_z$=+1 bright state \cite{Furdyna}. These effects suppress the exciton-Mn energy transfer rate $k_{Mn}$ and therefore suppress (enhance) the Mn$^{2+}$ (exciton) PL. In contrast, ``dark" ($J_z$=$\pm$2) excitons have $s^{ex}_z=\pm 1$ and \emph{can} excite the $S^{Mn}_z$=-5/2 ground state; this explains why, e.g., PL intensities in epitaxial dots are relatively unaffected in the Voigt geometry (\textbf{B}$\perp$\textbf{k}, \textbf{B}$\perp$ dot axis), where bright and dark excitons become mixed \cite{Chernenko}.

Once the Mn$^{2+}$ $3d^5$ shell is excited, relaxation to the $^4T_1$ level proceeds quickly \cite{Chernenko}. From this point, however, radiative transitions to the $^6A_1$ ground state are nominally spin-forbidden. However, weak spin-orbit couplings of the $^4T_1$ levels soften these selection rules \cite{Furdyna, Becker}, allowing phonon-assisted Mn$^{2+}$ PL at $\sim$2.1~eV.

In marked contrast to the conventional behavior just described, Figs. 1(c,d) show magneto-PL from Mn:ZnSe/CdSe nanocrystals.  At zero field, only the Mn$^{2+}$ PL band is visible -- exciton PL is entirely quenched (undoped but otherwise identical NCs \emph{do} show strong exciton PL at $\sim$2.7 eV; not shown). Thus, energy transfer from excitons to the Mn$^{2+}$ ions is quite efficient in these NCs even though the average Mn concentration is $<$1\%. More importantly, the Mn$^{2+}$ PL band is \emph{not} suppressed in applied magnetic fields -- rather, it develops a pronounced circular polarization of $\sim$30\%. To our knowledge, this behavior has no precedent in any nonresonant PL study of A$^{\textmd{II}}_{1-x}$Mn$_x$B$^{\textmd{VI}}$ materials reported to date \cite{Furdyna, Becker, LeeRamdas, Kim, Lee, Falk, Oh, Shibata, Hundt, Chernenko}. These data therefore provide a first indication that, in contrast to typical DMS materials, the Mn$^{2+}$ PL from DMS nanocrystals is more than merely a byproduct of exciton-Mn energy transfer.  Rather, the data point to the fundamentally different character of both the Mn$^{2+}$ excitation and emission process in strongly quantum-confined systems which, as elucidated below, can actually be used to reveal spin physics and \emph{sp-d} interactions in nanoscale DMS materials.

We note that these new findings are unrelated to the circularly-polarized \emph{exciton} PL reported in large Mn:CdSe NCs \cite{Beaulac}. By design, excitons in large Mn:CdSe NCs have energies $<$2.1~eV, so that energy transfer to the Mn$^{2+}$ does not occur. Thus, exciton PL remains unquenched and, in keeping with traditional DMS, becomes circularly polarized.

It is also evident in Fig. 1(c) that exciton PL at the NC band edge never appears in applied magnetic fields to 6~T (and even in ultrahigh fields to 55~T; not shown). Again, this stands in marked contrast to conventional DMSs \cite{Furdyna, Becker, LeeRamdas, Kim, Lee, Falk, Oh, Shibata, Hundt, Chernenko}, where band-edge PL increases dramatically with field because the exciton-Mn$^{2+}$ energy transfer rate $k_{Mn}$ is suppressed. Thus, exciton-Mn energy transfer remains surprisingly efficient in these NCs, \emph{even when} the Mn$^{2+}$ are completely polarized ($S^{Mn}_z$=-5/2). This contrast with conventional DMSs likely originates not only in the strong exciton-Mn coupling that exists in colloidal NCs, but also in the essential role played by dark excitons (which, as described above, can excite Mn ions regardless of $S_z^{Mn}$).  Specifically, i) dark excitons may already be the preferred ground state in applied fields since the \emph{s-d} exchange constant is likely inverted in these NCs \cite{Merkulov, Bussian}, ii) ground-state (i.e., dark) excitons in NCs have very small radiative rates at low temperatures ($k_R$$\ll$$k_{Mn}$) \cite{Crooker}, and iii) even if bright excitons have lower energy, the random NC orientation mixes bright with dark excitons since \textbf{B} will seldom lie along preferred angular momentum quantization axes.

\begin{figure}[tbp]
\includegraphics[width=.48\textwidth]{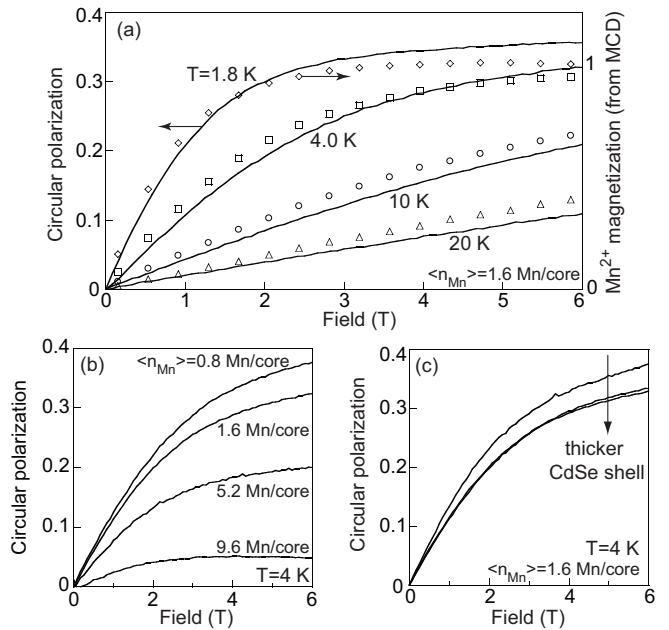}
\caption{(a) The circular polarization (\emph{CP}) of the Mn$^{2+}$ PL from Mn:ZnSe/CdSe nanocrystals versus field and temperature (lines; left axis). \emph{CP} tracks the Brillouin-like Mn$^{2+}$ magnetization (independently measured by MCD; points, right axis). (b) $CP(B)$ for different $\langle n_{Mn}\rangle$ (all NCs have similar CdSe shells). (c) $CP(B)$ for NCs with different CdSe shell thickness (all cores have $\langle n_{Mn}\rangle$=1.6).}
\label{fig2}
\end{figure}

Importantly, Figure 2 reveals that the circular polarization of the Mn$^{2+}$ PL follows the paramagnetic (Brillouin-like) magnetization of the spin-5/2 Mn$^{2+}$ ions. The circular polarization, $CP$=$(I^{\sigma+} - I^{\sigma-})/(I^{\sigma+} + I^{\sigma-})$ (solid line, left axis), scales with the temperature- and field-dependent Mn$^{2+}$ magnetization (symbols, right axis), which was independently measured in this same NC ensemble via MCD spectroscopy. Studies of NCs with different Mn doping [Fig. 2(b)] indicate that $CP$ is largest ($\sim$40\%) at the lowest Mn doping, but drops as $\langle n_{Mn}\rangle$ increases to 9.6 Mn/core, where Mn-Mn interactions become increasingly probable. Furthermore, Fig. 2(c) shows that for NCs with fixed $\langle n_{Mn}\rangle$ but increasing CdSe shell thickness, \emph{CP} is relatively unaffected.  Together, these data suggest that polarized Mn PL arises primarily from isolated Mn$^{2+}$ in the ZnSe cores, but is not particularly influenced by the overall NC size.

Perhaps most interestingly, and again in contrast to conventional DMS materials, we find that the Mn$^{2+}$ PL exhibits a sizable energy splitting, $\Delta E$, between $\sigma^\pm$ components [see Fig. 3(a)]. Fitting the uppermost 25$\%$ of the PL to a gaussian lineshape, we report here the center wavelength. Significantly, Fig. 3(b) shows that $\Delta E$ does \emph{not} increase linearly with field. Rather, $\Delta E$ exhibits a characteristic Brillouin-like dependence on field and temperature, that once again is proportional to the Mn$^{2+}$ magnetization within any given NC sample (subtracting field-independent PL backgrounds prior to fitting does not change $\Delta E$ appreciably). As such, $\Delta E$ saturates by only a few tesla at low temperatures -- but at large values of order 5-10~meV, which greatly exceeds the typical linear Zeeman shifts expected of Mn$^{2+}$ $3d^5$ levels ($g_{Mn}\simeq$2.0 $\equiv$ 60-300~$\mu$eV/T).

\begin{figure}[tbp]
\includegraphics[width=.48\textwidth]{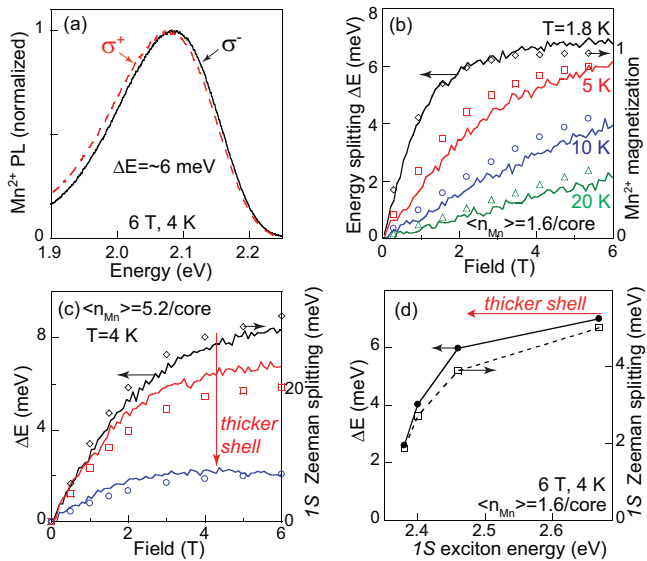}
\caption{(a) Normalized $\sigma^\pm$ Mn$^{2+}$ PL from Mn:ZnSe/CdSe NCs at 6~T, showing an energy splitting $\Delta E$. (b) $\Delta E$ (lines; left axis) scales with the Mn$^{2+}$ magnetization in this sample (points, right axis, from MCD). (c) Using NCs with identical Mn:ZnSe cores but increasing CdSe shell thickness to 6$\pm$2 \AA, $\Delta E$ tracks the \emph{1S} exciton Zeeman splitting (from MCD), indicating that $\Delta E$ scales with the net \emph{sp-d} coupling strength. (d) A similar comparison, in the $\langle n_{Mn} \rangle$=1.6 Mn/core NCs.} \label{fig3}
\end{figure}

Moreover, while $\Delta E$ is found to be largely independent of $\langle n_{Mn} \rangle$, it \emph{does} depend on the degree of quantum confinement in these core-shell NCs, which we can tune by varying the CdSe shell thickness. Figure 3(c) shows $\Delta E$ measured in a series of NCs with identical Mn:ZnSe cores, but with increasing CdSe shell thickness. While still following a Brillouin function, the saturation amplitude of $\Delta E$ is considerably reduced when the CdSe shell is thicker. Crucially, the reduction in $\Delta E$ is almost perfectly echoed by the similar reduction of the \emph{sp-d} coupling strength in the same three samples (as measured by MCD via the \emph{1S} Zeeman splitting). As shown previously \cite{Bussian}, thicker CdSe shells cause the conduction- and valence-band wavefunctions to occupy larger volumes, thereby reducing their overlap with the embedded Mn$^{2+}$ ions and reducing the overall \emph{sp-d} interaction strength -- in this way the \emph{sp-d} coupling can be independently tuned in these NCs while leaving the Mn doping unchanged. A similar agreement is observed in the other NC series; Fig. 3(d) compares the saturated (6~T) values of $\Delta E$ and \emph{1S} Zeeman splitting for the $\langle n_{Mn} \rangle$=1.6 series.  Again, both are reduced by the same factor as the \emph{sp-d} coupling is reduced by increasing CdSe shell thickness.  These data demonstrate that $\Delta E$ provides a reliable and independent measure of \emph{sp-d} coupling in strongly quantum confined DMS systems.

We note finally that neither \emph{CP} nor $\Delta E$ was found to be influenced by the polarization of the excitation light, even when resonantly pumping the \emph{1S} exciton peak with a tunable laser -- the Mn PL evidently retains little memory of the exciton's initial spin orientation.

We turn now to the possible origins of the Mn$^{2+}$ PL polarization and energy splitting.  Early studies of Mn:ZnS NCs revealed fast Mn PL decays \cite{Bhargava}, prompting suggestions that quantum confinement enhances hybridization between \emph{s,p} bands and Mn \emph{d} electrons. This point of view was later clarified \cite{Bol, Chen}: multi-exponential PL decays are a generic feature of A$^{\textmd{II}}_{1-x}$Mn$_x$B$^{\textmd{VI}}$ DMS regardless of dimensionality. We measured $\sim$100 $\mu$s PL decays in our NCs (independent of \textbf{B}), which is not atypical for dilute Mn$^{2+}$ in ZnSe.  Nonetheless, the circular polarization and (especially) the confinement-dependent $\Delta E$ both point to a coupling of the Mn $3d^5$ electrons with the NC's conduction and valence bands, beyond that typically found in more weakly-confined DMS materials. Related conclusions can also be drawn from the recent discovery that excitons can organize magnetic polarons in colloidal Mn:CdSe NCs, even at quite high temperatures \cite{BeaulacPolaron}.  While awaiting a formal theoretical underpinning of these effects, we anticipate that studies of single Mn-doped NCs will help unravel the nature of this coupling, as will measurements of ultrafast exciton dynamics as a function of NC size, \textbf{B}, and Mn$^{2+}$ doping. The role of any excess charge in these NCs, and how it may influence the polarization and energy of Mn$^{2+}$ emission, also merits investigation \cite{Efros}.

In summary, these studies reveal that the processes of both Mn$^{2+}$ excitation and Mn$^{2+}$ emission are essentially different in colloidal DMS nanocrystals as compared to conventional DMS materials.  Magnetic fields do not suppress the efficient exciton-Mn energy transfer, implicating the potential role of dark excitons in these NCs. Further, the circular Mn PL polarization and unexpectedly large energy splitting (that scales with \emph{sp-d} coupling strength) highlights the critical influence of strong quantum confinement and suggests their use as powerful probes of spin interactions in nanoscale DMS materials. This work was supported by the DOE Basic Energy Sciences Chem-, Bio- and Geosciences Division. We thank N. Samarth for the ZnMnSe epilayers, and J. Gaj, D. Yakovlev, and Al. Efros for valuable discussions.


\begin{references}

\bibitem{Furdyna}J. K. Furdyna, J. Appl. Phys. \textbf{64}, R29 (1988); O. Goede and W. Heimbrodt, Phys. Stat. Sol. B \textbf{146}, 11 (1988).

\bibitem{Becker}W. M. Becker, in \emph{Semiconductors and Semimetals, vol. 25 - Diluted Magnetic Semiconductors} (Eds: J. K. Furdyna, J. Kossut, Academic Press, San Diego 1988); T. Dietl, in \emph{Handbook on Semiconductors, Vol. 3b} (Eds: T.S. Moss, S. Mahajan, North-Holland, Amsterdam 1994).

\bibitem{Book}\emph{Intro. to the Physics of Diluted Magnetic Semiconductors} (Eds: J. Kossut, J. A. Gaj, Springer, Berlin 2010).

\bibitem{Hoffman}D. M. Hoffman \emph{et al.} Solid State Comm. \textbf{114}, 547 (2000).

\bibitem{Norris}D. J. Norris \emph{et al.}, Nano Letters \textbf{1}, 3 (2001).

\bibitem{Besombes}L. Besombes \emph{et al.}, Phys. Rev. Lett. \textbf{93}, 207403 (2004).

\bibitem{Hawrylak}R. M. Abolfath \emph{et al.}, Phys. Rev. Lett. \textbf{98}, 207203 (2007).

\bibitem{Merkulov}I. A. Merkulov \emph{et al.}, Phys. Rev. Lett. \textbf{83}, 1431 (1999).

\bibitem{Stern}N. P. Stern \emph{et al.}, Phys. Rev. B \textbf{75}, 045329 (2007).

\bibitem{Bussian}D. A. Bussian \emph{et al.}, Nat. Mat. \textbf{8}, 35 (2009).

\bibitem{LeeRamdas}Y. R. Lee \emph{et al.}, Phys. Rev. B \textbf{38}, 10600 (1988); J. F. MacKay \emph{et al.}, Phys. Rev. B \textbf{42}, 1743 (1990). 

\bibitem{Kim}C. S. Kim \emph{et al.}, J. Crystal Growth \textbf{214/215}, 395 (2000). 

\bibitem{Lee}S. Lee \emph{et al.}, Phys. Rev. B \textbf{72}, 075320 (2005). 


\bibitem{Falk}H. Falk \emph{et al.}, Phys. Rev. B \textbf{68}, 165203 (2003); V. F. Agekyan \emph{et al.}, Phys. Sol. State \textbf{52}, 27 (2010). 

\bibitem{Oh}E. Oh \emph{et al.}, Appl. Phys. Lett. \textbf{93}, 041911 (2008); B. J. Cooley \emph{et al.}, Nano Lett. \textbf{9}, 3142 (2009). 

\bibitem{Shibata}Y. Oka \emph{et al.}, J. Lumin. \textbf{100}, 175 (2002). 

\bibitem{Hundt}A. Hundt \emph{et al.}, Phys. Rev. B \textbf{69}, 121309(R) (2004). 

\bibitem{Chernenko}A. V. Chernenko \emph{et al.},  Phys. Rev. B \textbf{72}, 045302 (2005); A. V. Chernenko \emph{et al.}, J. Phys.:Cond. Matt. \textbf{22}, 355306 (2010).

\bibitem{Ratna}H. S. Yang \emph{et al.}, J. Appl. Phys. \textbf{93}, 586 (2003).

\bibitem{Beaulac}R. Beaulac \emph{et al.}, Adv. Funct. Mat. \textbf{18}, 3873 (2008).


\bibitem{Ivanov}S. Ivanov \emph{et al.}, J. Phys. Chem. B \textbf{108}, 10625 (2004).

\bibitem{Nawrocki}M. Nawrocki \emph{et al.}, Phys. Rev. B \textbf{52}, R2241 (1995); V. G. Abramishvili \emph{et al.}, Solid State Comm. \textbf{78}, 1069 (1991).

\bibitem{Crooker}S. A. Crooker \emph{et al.}, Appl. Phys. Lett. \textbf{82}, 2793 (2003).




\bibitem{Bhargava} R. N. Bhargava \emph{et al.}, Phys. Rev. Lett. \textbf{72}, 416 (1994).

\bibitem{Bol}A. A. Bol, A. Meijerink, Phys. Rev. B \textbf{58}, R15997 (1998); M. A. Chamarro \emph{et al.}, J. Crys. Growth \textbf{159}, 853 (1996). 

\bibitem{Chen}L. Chen \emph{et al.}, Phys. Stat. Solidi B \textbf{247}, 2522 (2010).

\bibitem{BeaulacPolaron}R. Beaulac \emph{et al.}, Science \textbf{325}, 973 (2009).

\bibitem{Efros} Al. L. Efros, private communication.



\end{references}
\end{document}